\documentclass[letterpaper,notoc]{JHEP3} 
\JHEPspecialurl{http://jhep.sissa.it/JOURNAL/JHEP3.tar.gz}
\usepackage{epsfig,multicol}
\newcommand{\met}{\mbox{$\rlap{\kern0.25em/}E_T$}}
\newcommand{\mpt}{\mbox{$\rlap{\kern0.25em/}p_T$}}
\title{\boldmath$pp\to t\bar{t}H\,,\,H\to\tau^+\tau^-$\unboldmath:
  toward a model independent determination of the Higgs boson
  couplings at the LHC}
\author{Alexander Belyaev and Laura Reina\\
  Florida State University, Tallahassee, FL 32306-4350, USA\\
  E-mail: \email{belyaev@hep.fsu.edu}, \email{reina@hep.fsu.edu}}
\preprint{\hepph{yymmxxx}}
\abstract{The possibility of detecting a Higgs boson through several
  production and decay channels is instrumental to the measurement of
  its couplings. In this paper we study the $pp\to t\bar
  tH,\,H\to\tau^+\tau^-$ channel at the LHC, for the case of a scalar
  Higgs boson, and use the obtained results to improve on existing
  strategies toward a model independent determination of the Higgs
  boson couplings. The case of a scalar Higgs boson with mass below
  140~GeV looks particularly promising.}
\keywords{Higgs boson, hadron colliders}
\begin{document}
\bibliographystyle{JHEP}
\section{Introduction}
\label{sec:intro}
The search for a Higgs boson in the mass region between the
experimental lower bound and the $Z$-boson production threshold is
among the most important goals of present and future hadron colliders.
The lower bound on the Higgs boson mass has been set by LEP at $M_H\ge
114.1$~GeV for a Standard Model (SM) Higgs boson \cite{Lepewwg:2001a},
and at $M_{h^0}\ge 91.0$~GeV and $M_{A^0}\ge 91.9$~GeV respectively
for the light scalar and the pseudoscalar Higgs bosons of the Minimal
Supersymmetric Model (MSSM) \cite{Lepewwg:2001b}.  At the same time,
precision fits of the Standard Model point to the existence of a SM
Higgs boson with mass below approximately 196~GeV~\cite{Lepewwg:2001},
while the MSSM requires the lightest scalar Higgs boson to have a mass
below approximately $135$~GeV~\cite{Heinemeyer:1998np}.

Evidence for a Higgs boson particle in the mass range up to about
180~GeV could be provided by the Run II of the Fermilab Tevatron. The
discovery of a relatively light Higgs boson at the Tevatron is in fact
one of the greatest expectations we can harbor for the pre-LHC era.
However, the statistics available at the Tevatron will not be enough
to measure the couplings of the discovered Higgs boson.  Higher energy
hadron colliders, like the LHC (Large Hadron Collider) and its
upgrade, the SLHC, or even higher energy future hadron colliders like
a VLHC (Very Large Hadron Collider), will explore the entire Higgs
boson mass spectrum up to the TeV scale, and will also have enough
statistics to shed some light on the pattern of its interactions. A
high energy $e^+e^-$ Linear Collider could then provide the best
environment to frame the nature of any discovered Higgs boson
unambiguously. In this context, the indications coming from the LHC
will be extremely important, and all efforts should be made to use the
potential of this machine at best.
 
During the last few years we have witnessed a dramatic improvement in
both experimental and theoretical studies of several Higgs boson
production channels and decay modes. In spite of the fact that
gluon-gluon fusion, $gg\to H$, is the leading scalar Higgs boson
production mode at the LHC and its upgrades, subleading production
modes, like weak boson fusion, $qq\to qqH$, and associated production
with top quark pairs, $pp\to t\bar tH$, are extremely important to
provide complementary informations and allow unique determinations of
ratios of Higgs boson couplings.  Strategies like the one proposed in
Ref.~\cite{Zeppenfeld:2000td} show how the combined informations from
both $gg\to H,\,(H\to
\gamma\gamma,ZZ,WW)$~\cite{TDRATLAS:1999,TDRCMS:1994,Denegri:2001pn}
and $qq\to qqH,\,(H\to
\gamma\gamma,\tau\tau,WW)$~\cite{Rainwater:1997dg,Rainwater:1999gg,
  Rainwater:1998kj,Plehn:1999xi, Rainwater:1999sd,Kauer:2000hi} can
confirm the Standard Model (SM) paradigm and, under this assumption,
determine the width of the discovered Higgs boson with good accuracy.
These studies have been further updated to include the results of
recent analyses of $pp\to t\bar tH,\,(H\to b\bar b,\, W^+W^-)$
\cite{Richter-Was:1999sa,Beneke:2000hk,Drollinger:2001ym,Maltoni:2002jr}
and preliminary results on $pp\to WH,\, H\to b\bar b$
\cite{Drollinger:2002uj}, and estimates of the accuracies on
individual SM Higgs boson couplings to both SM fermions and gauge
bosons have been presented
\cite{Zeppenfeld:2002ng,Cavalli:2002vs,Conway:2002kk}.

In this paper we study the $pp\to t\bar tH,\, H\to\tau^+\tau^-$
channel, where a scalar Higgs boson $H$ is radiated off a top quark or
antiquark and decays into a $\tau^+\tau^-$ pair, at the LHC with
$\sqrt{s}\!=\!14$~TeV, and we use our results to improve on existing
strategies toward a model independent determination of the couplings
of a scalar Higgs boson to both SM fermions and gauge bosons. The
scalar Higgs boson $H$ can be thought as the Higgs boson of the SM or
the light Higgs boson of a Two Higgs Doublet Model (2HDM), including
the MSSM, or a light scalar Higgs boson of any other more general
extension of the minimal scalar sector of the SM. Indeed, studying the
properties of a light scalar Higgs boson can be instrumental in
disentangling the first evidence of new physics beyond the
SM\footnote{For the case of a generic new scalar particle, not
  necessarily a Higgs boson, see e.g.~Ref.~\cite{Burgess:1999ha}}.

In particular, we find that a scalar Higgs boson can be observed with
good accuracy in the $pp\to t\bar{t}H,\, H\to\tau^+\tau^-$ channel for
masses $M_H\!\le\!140$~GeV.  For Higgs masses $M_H\!\le\!140$~GeV we
then have the possibility of measuring both $pp\to t\bar{t}H,\,H\to
b\bar b$ and $pp\to t\bar{t}H,\, H\to\tau^+\tau^-$.  This allows us to
remove the assumption, made in all existing analyses, that the ratio
of the bottom quark Yukawa coupling ($y_b$) to the $\tau$ lepton
Yukawa coupling ($y_\tau$) is SM-like, i.e. it goes as the ratio of
the corresponding masses.  Although many models beyond the Standard
Model are compatible with this assumption at tree level, they can show
sizable deviations due to non SM-like loop corrections, which would be
missed if the SM-likeness of this ratio had to be assumed. This is,
for instance, the case of the MSSM in certain region of its parameter
space
\cite{Hall:1994gn,Carena:1994bv,Carena:1999py,Borzumati:1999sp,Carena:2001bg,Guasch:2001wv}.
We note that the ratio $y_b/y_\tau$ could also be determined by
combining $qq\to qqH,\, H\to\tau^+\tau^-$ and $q\bar{q}\to WH,\, H\to
b\bar b$, where however very high luminosities are required to obtain
acceptable accuracies in the $WH$ channel. With this respect, the
possibility to determine the ratio $y_b/y_\tau$ via the $pp\to t\bar t
H$ production channel alone represents a definite improvement. Indeed,
this measurement is free of the ambiguity that comes from the
experimental inability of distinguish between $WWH$ and $ZZH$ weak
boson fusion in $qq\to qqH$, and can be obtained with lower
luminosities.

We also explicitly release the assumption that the $Hgg$ loop-induced
coupling ($y_g$) is mainly determined by the contribution of the top
quark loop, as in the SM, and is therefore proportional to the top
quark Yukawa coupling ($y_t$). We treat $y_g$ and $y_t$ as independent
couplings. This is particularly important in view of recent studies
that point at deviations of the $Hgg$ coupling from the SM-paradigm in
models with extra dimensions
\cite{Giudice:2000av,Hewett:2002nk,Cacciapaglia:2001nz}.

We propose a general strategy to determine the width and couplings of
a scalar Higgs boson to the SM fermions and gauge bosons which is
model independent and could be applied to any scenario of new physics
beyond the SM. As a numerical example, we specify it to the case of a
SM-like Higgs boson, since most existing experimental studies have
been performed under this assumption. It is moreover reasonable to
expect that the experimental accuracies determined in these studies
apply also to the case of a generic scalar Higgs boson whose
properties do not differ dramatically from the SM Higgs boson.  Big
deviations from the Standard Model pattern, like a large enhancement
or suppression of certain production or decay modes, will anyhow
manifest themselves independently of any precision study of Higgs
physics.  In particular, we consider an intermediate mass scalar Higgs
boson, and we distinguish between two main mass regions, for Higgs
boson masses below and above 140~GeV. For $M_H\!\le\!140$~GeV we add
to the existing studies the $pp\to t\bar tH,\, H\to\tau^+\tau^-$
channel studied in this paper and we determine the accuracy expected
on the determination of the width, and on the determination of ratios
of couplings and individual couplings of the scalar Higgs boson to the
SM fermions and gauge bosons. For $M_H\!>\!140$~GeV both the $H\to
b\bar b$ and the $H\to\tau^+\tau^-$ branching ratios are very
suppressed and the accuracy of the corresponding channels becomes
problematic.  However, as pointed out in
Refs.~\cite{Zeppenfeld:2002ng,Maltoni:2002jr}, in this region we can
still focus on ratios of Higgs boson couplings and determine them with
good precision.

Assuming that the invisible component of the Higgs boson width is
negligible, very likely new heavy degrees of freedom will reveal
themselves as small deviations in the loop-induced couplings $Hgg$ and
$H\gamma\gamma$.  We therefore parameterize these couplings in terms
of the loop contribution of the SM particles plus a contribution due
to non SM heavy degrees of freedom.  As an outcome of our analysis we
will be able to estimate the sensitivity of the LHC to deviations of
the $Hgg$ and $H\gamma\gamma$ couplings due to non SM heavy degrees of
freedom.

The outline of our paper is as follows.  In Section~\ref{sec:tautau}
we present the study of the $pp\to t\bar tH,\, H\to\tau^+\tau^-$
channel, while in Section~\ref{sec:higgs_couplings} we describe the
strategy we adopt to determine the couplings of a light scalar Higgs
boson at the LHC. We summarize and present our conclusions in
Section~\ref{sec:conclusions}.

\boldmath
\section{A study of the $pp\to t\bar tH$ channel with $H\to\tau^+\tau^-$}
\label{sec:tautau}
\unboldmath

In this section we study the potential of the LHC to measure the cross
section for the $pp\to t\bar{t}H,\,H\to\tau^+\tau^-$ process, where
$H$ is a SM-like scalar Higgs boson. This measures the product of the
top quark and $\tau$ lepton Yukawa couplings $y_t\times y_\tau$ and,
as we will show in Section~\ref{sec:higgs_couplings}, is instrumental
to the model-independent measurement of the couplings of a light
scalar Higgs boson to both SM fermions and gauge bosons. 

We assume the LHC will run at $\sqrt{s}\!=\!14$~Tev with a total
integrated luminosity of $10^{34}$~cm$^{-2}$s$^{-1}$.  Both signal and
background are calculated at tree level in QCD, using the CompHEP
package~\cite{Pukhov:1999gg}. Accordingly, we use the CTEQ5L set of
parton distribution functions (PDF)~\cite{Lai:1999wy}, with
$\alpha_s(M_Z)\!=\!0.127$ at leading order of QCD. Both
renormalization and factorization scales have been set equal to the
invariant mass of the $\tau^+\tau^-$ pair: $\mu\!=\!m_{\tau\tau}$.
For the case of the signal $\mu\!=\!m_{\tau\tau}\!=\!M_H$. We note
that for this value of $\mu$, the tree-level signal cross section
effectively matches the NLO result~\cite{Beenakker:2001rj}, in the
intermediate Higgs boson mass region. Moreover, we use
$m_t\!=\!175$~GeV and $m_\tau\!=\!1.77$~GeV.  We have calculated the
Higgs boson width and the $H\!\to\!\tau^+\tau^-$ branching ratio using
the HDECAY program~\cite{Spira:1997if,Djouadi:1998yw}.

In order to unambiguously reconstruct both top quarks as well as the
Higgs boson mass, we choose the case when one of the top quarks decays
leptonically: $t\to b \ell\nu$ (where $\ell$ stands for electron or
muon and $\nu$ for the corresponding neutrino), and the other top
quark decays hadronically: $t\to b q q'$.  We only consider decays of
$\tau$-leptons into one or three charged pions. This gives one or
three charged tracks in the hadronic calorimeter, which, when
combined, form pencil-like narrow $\tau$-jets that we denote by
$\tau_j$.  We do not consider leptonic decays of $\tau$-leptons in
order to be able to reconstruct $two$ top quarks.  Therefore, for the
signal we study the $b\bar{b} \ell\nu q q' \tau^+\tau^-$ parton level
signature.

For the signal signature under study the only serious background is
the irreducible $t\bar{t}\tau^+\tau^-$ background, where the
$\tau^+\tau^-$ pair originates from a $Z$ boson or a photon. This
background has the same parton level and detector level signature as
the signal.  One can estimate the cross section of other possible
irreducible backgrounds, but they are all negligible compared to
$t\bar{t}\tau^+\tau^-$.  Indeed, the next obvious irreducible
background is $Wb\bar{b} j j \tau^+\tau^-$ (where $j$ stands for quark
or gluon). Even without considering the suppression due to
$Br(Z\to\tau^+\tau^-)$, its cross section is at least four orders of
magnitude lower then the cross section for $Wb\bar{b}$, because of an
additional $\alpha\alpha_s^2$ factor.  The cross section for the
$Wb\bar{b}$ process at the LHC is about 30~pb for $p_T^b>20$~GeV and a
$b$-quark separation cut $\Delta_R^{bb}>$0.5 ~\cite{Belyaev:1998dn}.
Therefore we estimate the cross section for $Wb\bar{b} j j
\tau^+\tau^-$ to be at most a few fb. In view also of the fact that we
reconstruct both top quarks, the contribution from this background
process can be safely neglected.  One can also think of reducible
backgrounds like $t\bar{t}b\bar{b}$ and $t\bar{t}jj$, as well as
$Wbjjj \tau^+\tau^-$, $Wjjjj \tau^+\tau^-$, $Wb\bar{b}jjW(W\to \tau
\nu)j$ or $Wb\bar{b}jjjj$, for which one or two jets are misidentified
as b-quarks or $\tau$-leptons. Taking into account that the
corresponding misidentification probabilities, 
$\epsilon_b$ is of the order of 1\% and $\epsilon_\tau$ is  the order
of 0.5\%, we find that these backgrounds represent at most a few
percent of the main $t\bar{t}\tau^+\tau^-$ background, when some of the
selection cuts described in the following are applied.

Therefore, for both signal and background we study $t\bar
t\tau^+\tau^-$ with a $b\bar{b} \ell\nu q q' \tau^+\tau^-$ parton
level signature, equivalent to a ($2b$-jets+jets+$\met$+2-$\tau_j$)
signature in the detector.  Moreover, for the $pp\to
t\bar{t}\tau^+\tau^-$ background process we require that
$m_{\tau\tau}\!>\!40$~GeV. This avoids considering contributions
irrelevant to our process. The numerical values of the cross sections
for both signal and $t\bar t\tau^+\tau^-$ background at the LHC with
$\sqrt{s}\!=\!14$~TeV are presented in Table~\ref{tab:cs_tttata}.
\TABLE[htb]{
\begin{tabular}{ |l||l|l|l|l|l|  }
\hline\hline
 & \multicolumn{1}{c|}{Background:}
 & \multicolumn{4}{c|}{Signal: $pp\to t\bar{t}H,\,H\to\tau^+\tau^-$} \\
 \cline{3-6}
 & \multicolumn{1}{c|}{$pp\to t\bar{t}\tau^+\tau^-$} 
 & 110 GeV & 120 GeV & 130 GeV & 140 GeV\\
  \hline
 CS(fb)  &  28.2   &  67.7  & 47.4  & 29.1   &  15.2  \\
  \hline
  \hline
 \end{tabular}
 \caption{\label{tab:cs_tttata} Cross sections for both signal, 
   $pp\to t\bar{t}H,\,H\to\tau^+\tau^-$, and background, $pp\to
   t\bar{t}\tau^+\tau^-$, processes at the LHC with
   $\sqrt{s}\!=\!14$~TeV.}}

For a realistic signal and background simulation we use the following
approach.
\begin{itemize}
\item Signal and background $t\bar{t}\tau^+\tau^- $  processes
      are calculated using the CompHEP~v4.1 package.
\item We use the CompHEP-PYTHIA interface~\cite{Belyaev:2000wn} to
      simulate the $b\bar{b} \ell\nu q q' \tau^+\tau^-$ signature.
\item We use TAUOLA~v2.6~\cite{Jadach:1993hs} to decay
      polarized $\tau$-leptons properly.  One should
      note that for the case of the background, the photon or Z-boson decay
      into $\tau_L^-\tau_R^+$ or $\tau_R^-\tau_L^+$ , whereas the scalar
      Higgs boson decays into either $\tau_L^- \tau_L^+$ or $\tau_R^-
      \tau_R^+$.  Taking into account the proper polarization of
      $\tau$-leptons is important, since pion energy distributions are
      just opposite in case of left- and right-polarized tau leptons
      (for details, see e.g.~\cite{Bullock:1993yt}).
\item  Although the analysis is done at the parton level,
        we take into account the detector energy resolution
        and apply  electron and jet energy Gaussian smearing
        for the electromagnetic and hadronic calorimeters, respectively:\\
        $\Delta E^{ele}/E=0.2/\sqrt{E}$ and $\Delta E^{had}/E=0.8/\sqrt{E}$.
\item To reproduce the realistic
        acceptance for leptons and quarks we require (CUT I):\\
        1) $p_T^\ell > 20$~GeV, $|\eta^\ell|<2.5$,
        $\Delta R(\ell,q)>0.4$,\\
        2) $p_T^b > 20$~GeV, $|\eta^b|<2$, $\Delta R>0.5$,\\ 
        as well as a $b$-quarks tagging efficiency of 60\%,\\
        3) $p_T^q > 20$~GeV, $|\eta^q|<3$,
        $\Delta R>0.5$,\\
        where $\Delta R$ is the $\delta\phi$ and $\delta\eta$
        separation: $\Delta R=\sqrt{\delta\phi^2+ \delta\eta^2}$, for
        $\phi$ the azimuthal angle and $\eta$ the pseudorapidity of
        a given particle.
\item To simulate the effective $\tau$-lepton identification
        (ID) tagging efficiency we require (CUT II):\\
        1) one or three charged $\pi$-mesons from $\tau$ decay,\\
        2) a cut on the minimum transverse momenta of each prong: 
           $p_T^\pi>5$~GeV,\\
        3) a cut on the total transverse momenta of $\tau_j$:     
           $p_T^{\tau_j}>20$~GeV, \\
        4) a cut on the pseudorapidity of $\tau_j$: $|\eta_{\tau_j}|<2$,\\
        5) $\Delta R_{\tau_j} >0.5$.\\
        The efficiency of these $\tau$-lepton ID cuts varies from
        $51\%$ to $58\%$: the lowest efficiency is for $\tau$ leptons
        from background while the highest efficiency is
        obtained for $M_H\!=\!140$~GeV signal events.
      \item We then follow the standard procedure of reconstructing
        the neutrino momentum from the leptonic decay of one of the
        top-quarks. We solve the equation for electron and missing
        transverse momenta to form the W-mass, and out of the two
        solutions for $p_Z^\nu$, we choose the solution having the
        absolute value of $|p_Z^\nu|$ which would be the right one in
        about $70\%$ of the cases.  The smearing of the total missing
        transverse momentum is simulated by calculating the missing
        momentum in the transverse plane, when all particles momenta
        are summed after having been properly smeared in the
        respective calorimeters, according to the calorimeter energy
        smearings discussed above. We have checked that our resolution
        for the missing transverse momentum is in a good agreement
        with the ATLAS TDR studies (see Figure 9-34
        in~\cite{TDRATLAS:1999}). One should also notice that we
        cannot distinguish between missing transverse momentum from
        $W$ and $\tau$ decays. This fact leads to the widening of
        top-quark invariant mass reconstructed in the leptonic
        channel.
        \item  For reconstructed top-quarks from the leptonic channel we
        require that their mass satisfies (CUT III):\\
        $175-m_t^\ell < 50$~GeV,\\ while for hadronically decaying top
        quarks we require their mass to be in the $\pm 50$~GeV mass window:\\
        $|m_t^h-175|<50$~GeV.
      \item Finally we reconstruct the Higgs boson mass by taking the
        invariant mass of two $\tau_j$:
        $M_H^{\tau\tau}=m_{\tau_j\tau_j}$. This variable should be
        considered as the effective mass of the Higgs boson since we
        do not take into account nor reconstruct the neutrinos from
        $\tau$-leptons decay.
\end{itemize}    

By following the strategy outlined above and assuming 100~fb$^{-1}$ of
total integrated luminosity (one detector), we obtain the overall
efficiencies for the $b\bar{b} \ell\nu q q' \tau^+\tau^-$ signature
shown in Table~\ref{tab:res_tttata}. In Figure~\ref{fig:mh_tttata} we
plot the corresponding number of events for both signal and background
as a function of the reconstructed invariant mass $M_H^{\tau\tau}$.
One can see that, for each mass in the range
$M_H\!=\!110\!-\!140$~GeV, the $M_H^{\tau\tau}$ distribution can be
fitted and direct correspondence to the real $M_H$ mass can be
established.  For $M_H\!=\!110\!-\!140$~GeV the expected accuracy of
the cross section measurement is the inverse significance,
$\delta\sigma/\sigma =\sqrt{S+B}/S$, shown in the last line of
Table~\ref{tab:res_tttata}.  The accuracy of the measurement of the
product of Yukawa couplings $y_t\times y_\tau$ is quite good and is
equal to half of the $\sqrt{S+B}/S$. It varies from 10 to 25\% for
$M_H\!=\!110\!-\!140$~GeV.
\TABLE[ht]{
\begin{tabular}{ |l||l|l|l|l|l|  }
\hline\hline
 & \multicolumn{1}{c|}{Background: }
 & \multicolumn{4}{c|}{Signal: $pp\to t\bar{t} H,\,H\to\tau^+\tau^-$} \\
 \cline{3-6}
 & \multicolumn{1}{c|}{$pp\to t\bar{t}\tau^+\tau^-$} 
 & 110 GeV & 120 GeV & 130 GeV & 140 GeV\\
  \hline
 Eff. of CUTS I+II+III ($\%$)   & 0.42 &  0.50 & 0.52 & 0.55  & 0.58   \\
 Number of events/100~fb$^{-1}$ & 12   &  34   & 25   & 16    & 8.8    \\
 $S/\sqrt{S+B}$                 &      &  5.0  & 4.1  & 3.0   & 1.9    \\
 $S/B$                          &      &  2.8  & 2.1  & 1.3   &  0.7   \\
 $\delta\sigma/\sigma$          &      &  0.20 & 0.24 & 0.33  &  0.52  \\
 \hline
  \hline
 \end{tabular}
 \caption{\label{tab:res_tttata} 
   Efficiency of CUTS I+II+III (see text) for both signal and
   background $b\bar{b} \ell\nu q q' \tau^+\tau^-$ signatures,
   together with the significance and precision of the the signal
   cross section measurements for $M_H\!=\!110,120,130$ and 140~GeV,
   at the LHC with $\sqrt{s}\!=\!14$~TeV. A total integrated
   luminosity of 100~fb$^{-1}$ is assumed}}
\FIGURE[ht]{
\noindent
\resizebox{!}{8cm}{\includegraphics{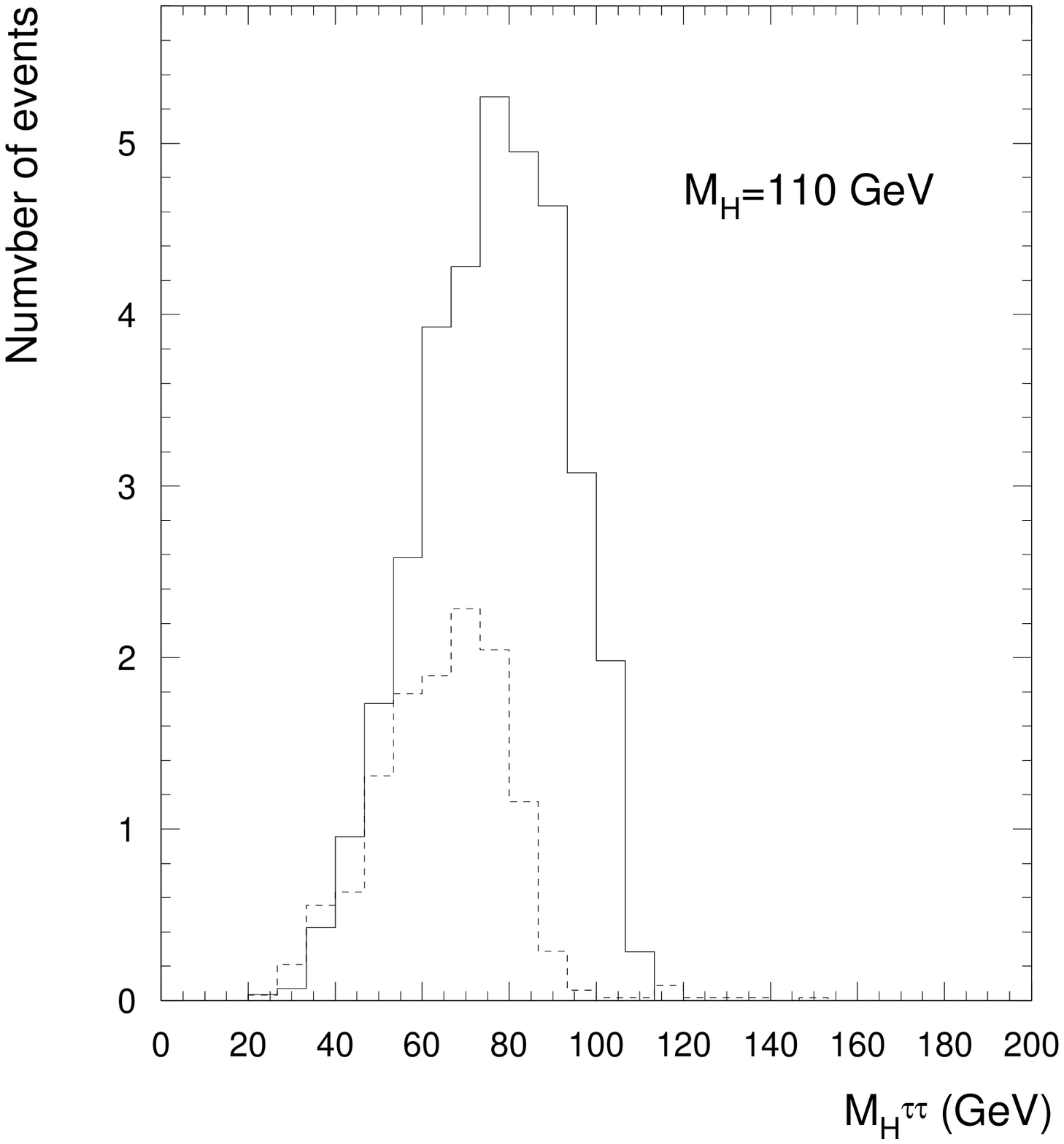}}
\resizebox{!}{8cm}{\includegraphics{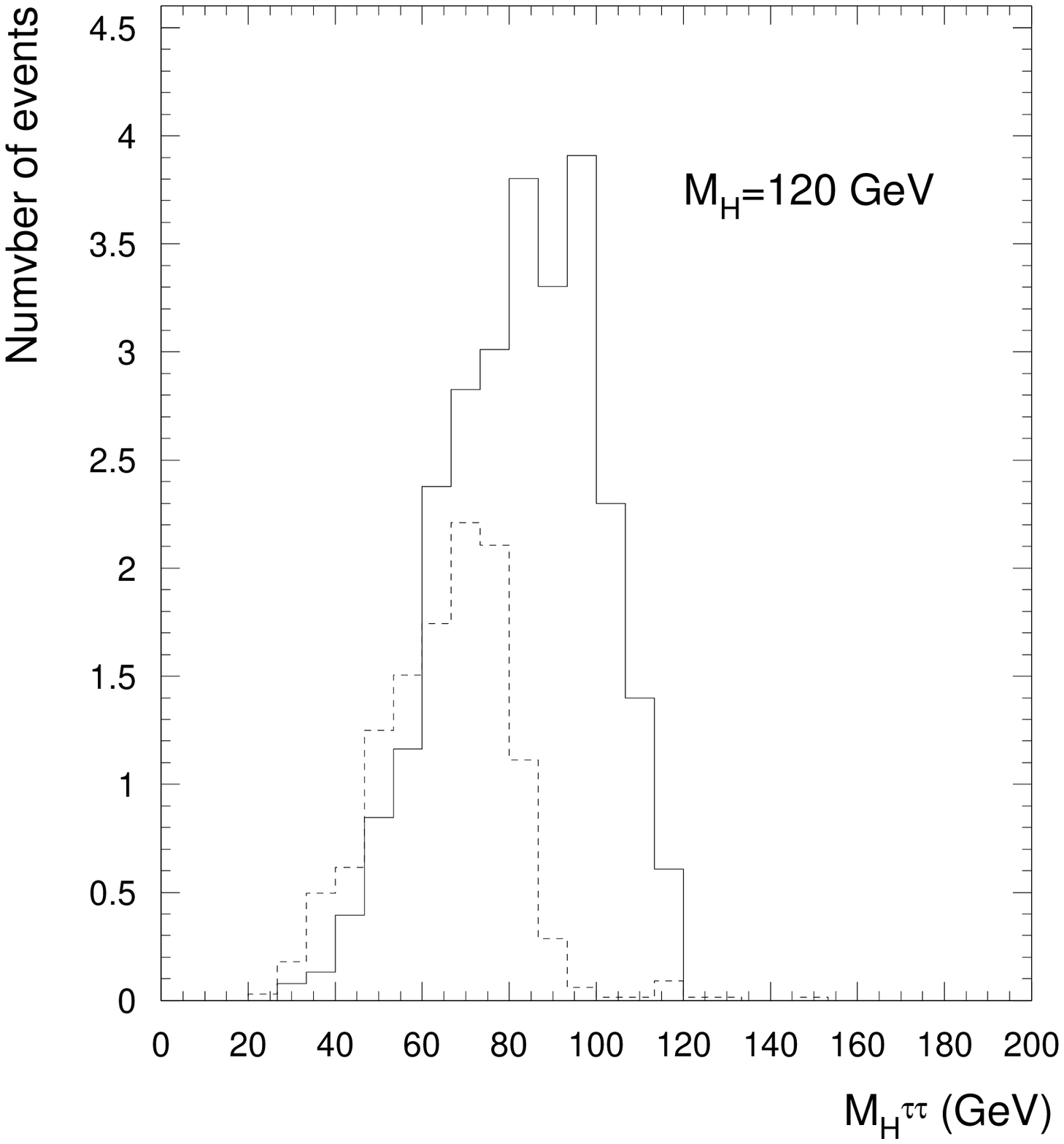}}\\
\resizebox{!}{8cm}{\includegraphics{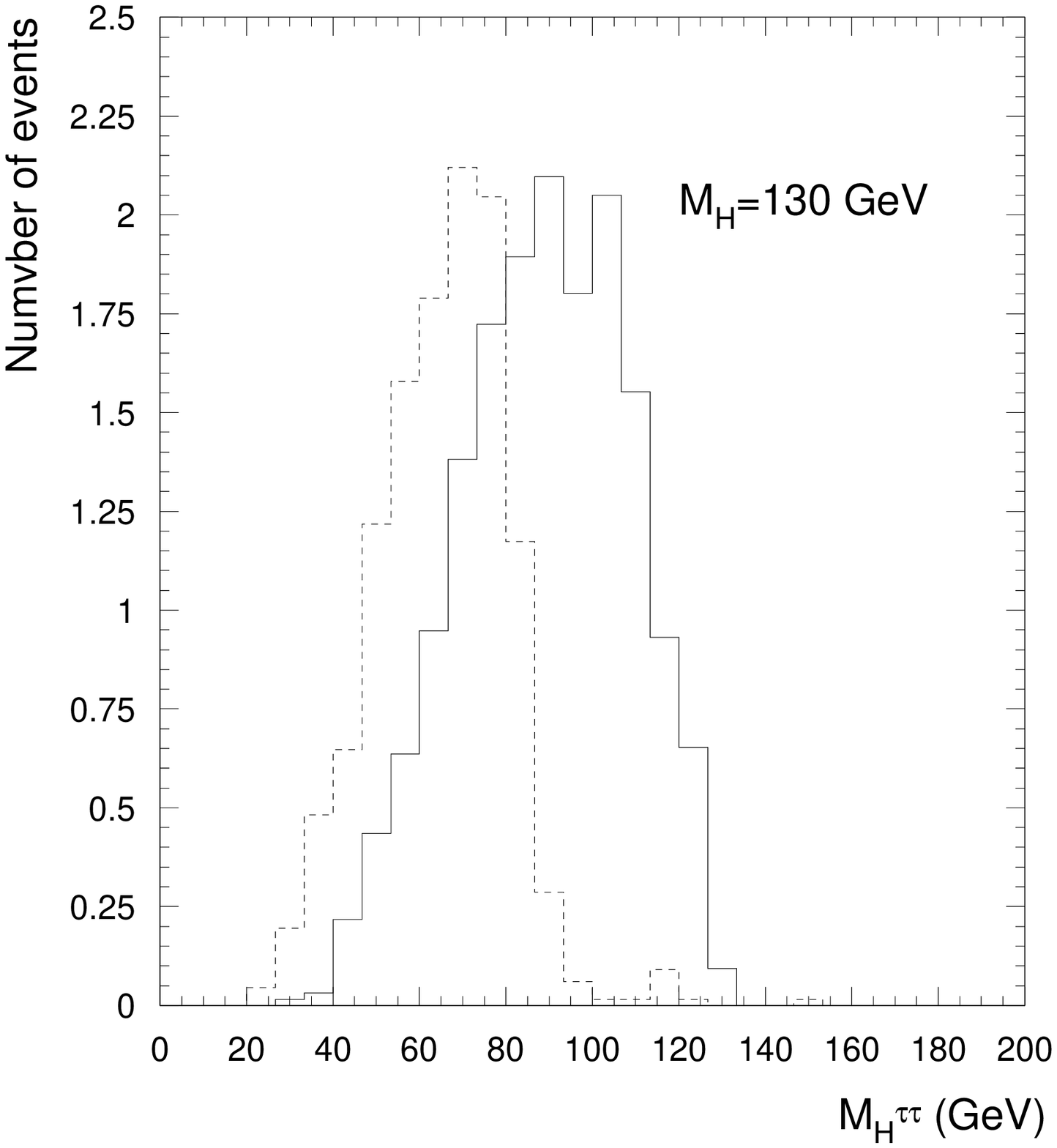}}
\resizebox{!}{8cm}{\includegraphics{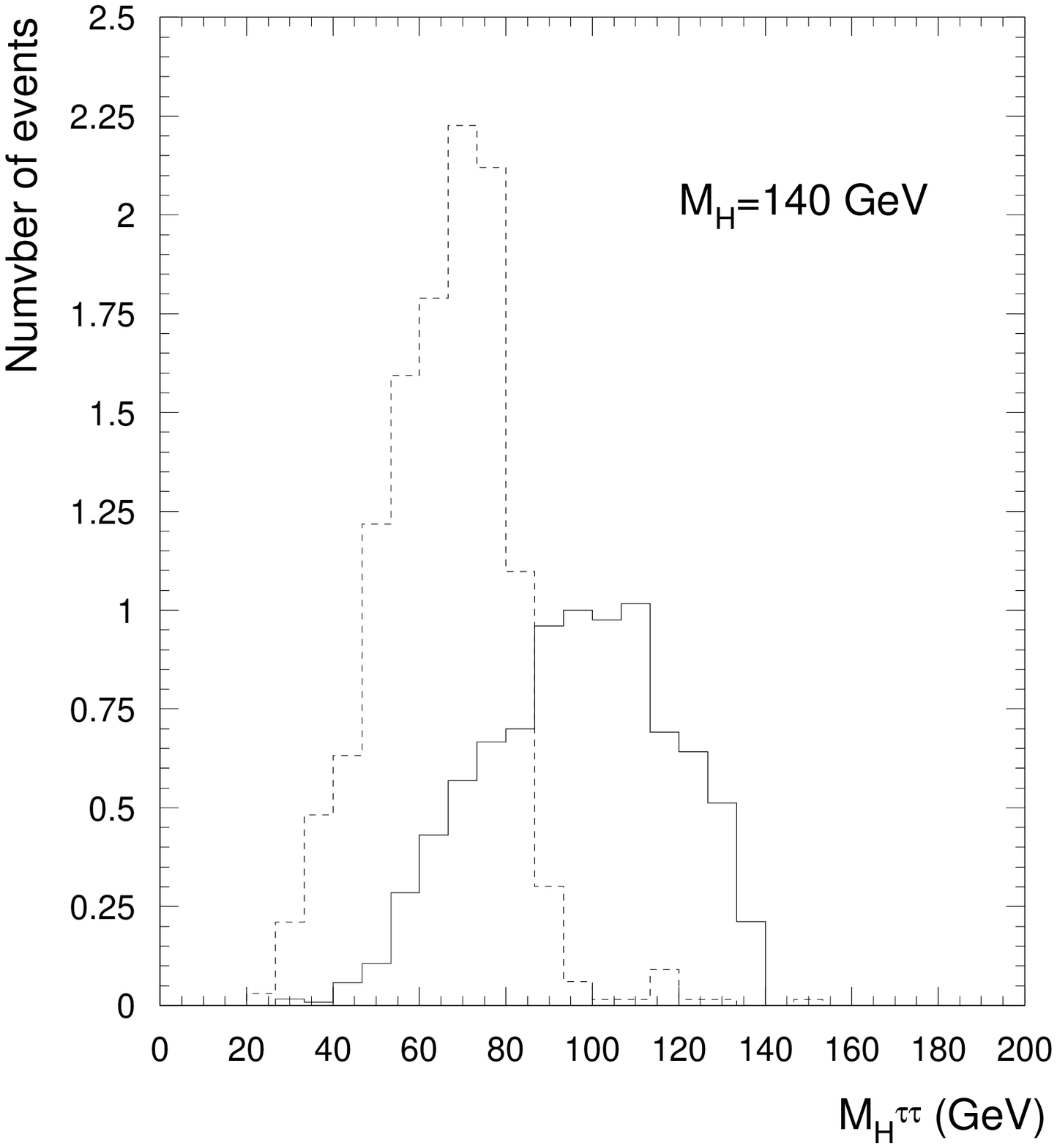}}
\caption{\label{fig:mh_tttata} The $M_H^{\tau\tau}$ invariant mass
  distributions for the signal (solid line) and background (dashed
  line) $t\bar{t}\tau^+\tau^-$ signature at the LHC with
  $\sqrt{s}\!=\!14$~TeV. A total integrated luminosity of
  100~fb$^{-1}$ is assumed.}}

\section{Determining the Higgs couplings}
\label{sec:higgs_couplings}

If a Higgs boson candidate is discovered with properties that differ
substantially from the SM Higgs, other more dramatically new processes
will be observed independently of any precision study of its
couplings. Some decay or production channels will be anomalously
enhanced or completely missing, and this will be enough to give strong
indications of the non-standard nature of the discovered particle.  If
this is not the case, a Higgs boson candidate will very likely have
production cross sections and decay branching ratios of roughly the
same order of magnitude of the SM Higgs boson.  Indications of its non
standard nature could come from precise measurements of individual
tree level ($y_i$ for $i=b,t,\tau,W,Z$) or loop induced ($y_g$ and
$y_\gamma$) couplings, and from the study of ratios of couplings, like
$y_b/y_\tau$ or $y_W/y_Z$, possibly obtained through different
production and decay channels.  Following the notation introduced in
\cite{Zeppenfeld:2000td}, we will develop our analysis in terms of
Higgs decay rates, by substituting each Higgs coupling square $y_i^2$
with the corresponding Higgs decay rate $\Gamma_i=\Gamma(H\to ii)$,
where the final state particles can be either real or virtual.  Ratios
of couplings can be then measured through the ratios of the
corresponding Higgs rates or branching ratios, and are particularly
interesting, as pointed out in Ref.~\cite{Zeppenfeld:2000td}, because
their theoretical predictions are less sensitive to QCD uncertainties,
PDF's uncertainties, and all sort of dependencies that equally affect
the channels that enter the ratio. Individual couplings can also be
measured, but their extraction requires some more elaborated strategy,
as we will explain in the following.

In our analysis we consider some of the production+decay channels that
have been studied in the literature for a SM-like Higgs boson and we
include the results for the $pp\to t\bar{t}H,\,H\to\tau^+\tau^-$
channel presented in Section~\ref{sec:tautau}. For completeness, we
summarize all the available channels in the following, together with
the main references where they have been studied:
\begin{eqnarray}
\label{eq:all_channels}
&&gg\rightarrow H\,\,\,
\mbox{with}\,\,\,H\rightarrow\gamma\gamma,\,ZZ,\,WW\,\,\,\,\,\,\,\,\,
\cite{TDRATLAS:1999,TDRCMS:1994,Denegri:2001pn}\,\,\,,\\
&&qq\rightarrow qq H\,\,\,\mbox{with}\,\,\,
H\rightarrow\gamma\gamma,\,\tau\tau,\,WW \,\,\,\,\,\,\,\,\,
\cite{Rainwater:1997dg,Rainwater:1999gg,Rainwater:1998kj,Plehn:1999xi,
Rainwater:1999sd,Kauer:2000hi}
\,\,\,,\nonumber\\
&&q\bar q,gg\rightarrow t\bar tH\,\,\,\mbox{with}\,\,\,
H\rightarrow b\bar b,\tau\tau,WW\,\,\,\,\,\,\,\,\,
\cite{Richter-Was:1999sa,Beneke:2000hk,Drollinger:2001ym,Maltoni:2002jr}\,\,
\mbox{and Section~\ref{sec:tautau}}
\,\,\,,\nonumber\\
&&q\bar q\rightarrow WH\,\,\,\mbox{with}\,\,\,H\rightarrow b\bar b
\,\,\,\,\,\,\,\,\,\cite{Drollinger:2002uj}\,\,\,.\nonumber
\end{eqnarray}
Each process in Eq.~(\ref{eq:all_channels}) depends on two Higgs
couplings, one from the Higgs boson production and one from the Higgs
boson decay, except the $qq\to qqH$ channels, that are actually
combinations of both $WWH$ and $ZZH$ fusion processes. These two modes
cannot be distinguished experimentally and the accuracies given in
Table~\ref{tab:accuracies} refers to the superposition of both.  To be
completely model independent one should work with the actual linear
superposition of both channels, where the couplings of both the $Z^0$
and the $W^\pm$ gauge bosons to fermions are well known, while the
$y_z$ and $y_w$ couplings are unknown. In so doing the analytical
expressions that we will present later on would become much less
transparent, and we therefore decide to keep the assumption that the
ratio between $y_z$ and $y_w$ is SM-like, and to remove only the model
dependence of the $y_b$ and $y_\tau$ couplings.  Since the couplings
of a scalar Higgs boson both to the $Z^0$ and the $W^\pm$ gauge bosons
are closely related to the EW SU(2) gauge symmetry, and since the EW
gauge interactions have been proven to be so successfully described by
the SM, it seems reasonable to assume that
$y_z/y_w\!=\!y_z^{SM}/y_w^{SM}$, or $\Gamma_z/\Gamma_w\!=\!z_{SM}$. We
note that for $M_H\!\ge\!130-140$ the assumption that
$\Gamma_z/\Gamma_w\!=\!z_{SM}$ can be directly tested at the 20-30\%
level by measuring $Z^{(g)}_w/Z^{(g)}_z$~\cite{Zeppenfeld:2002ng}.  As
we will discuss more in detail later on in this Section, thanks to the
availability of the $pp\to t\bar{t}H,\, H\to\tau^+\tau^-$ channel
studied in this paper this assumption can also be tested for
$M_H\!\le\!  130-140$~GeV, when integrated luminosities of the order
of 300~fb$^{-1}$ are available.

Working under the assumption that $\Gamma_z/\Gamma_w=z_{SM}$, the
observation of a scalar Higgs boson in any of the channels listed in
Eq.~(\ref{eq:all_channels}) provides a measurement of the
corresponding ratio $Z^{(i)}_j$ defined as
\begin{equation}
\label{eq:z_i_j}
Z^{(i)}_j=\frac{\Gamma_i\Gamma_j}{\Gamma}\,\,\,,
\end{equation}
where the apex $i\!=\!g,w,t$ indicates the production process, the
index $j\!=\!b,\tau,w,z,g,\gamma$ indicates the decay process, and, as
before, we have denoted by $\Gamma_j$ the decay rate for $H\to jj$ and
by $\Gamma$ the total Higgs boson width. Only exception to our
notation, for $q\bar q\to WH,\,H\to b\bar b$ we define
$Z^{(wH)}_b\!=\!\Gamma_w\Gamma_b/\Gamma$. We summarize in
Table~\ref{tab:accuracies} the relative accuracy estimated for each
$Z^{(i)}_j$ in the corresponding studies. We also illustrate these
accuracies in Fig.~\ref{fig:all_channels}, including the results on
$pp\to t\bar tH,\,H\to\tau^+\tau^-$ presented in
Section~\ref{sec:tautau} of this paper.  
\TABLE[h]{
\caption{Summary of the accuracies (\%) on the available production+decay
  channels for a SM-like scalar Higgs boson. All of them assume
  $2\times 100$~fb$^{-1}$ of integrated luminosity, i.e.
  $100$~fb$^{-1}$ of integrated luminosity per detector, except for
  $Z_w^{(w)}$ (30 fb$^{-1}$), $Z_b^{(wH)}$ (300 fb$^{-1}$), and
  $Z_w^{(t)}$ (300 fb$^{-1}$). The original values are from the
  references listed in Eq.~(\ref{eq:all_channels}) and the study
  presented in Section~\ref{sec:tautau} of this paper.}
\label{tab:accuracies}
\begin{tabular}{|c||c|c|c|c|c|c|c|c|}
\hline\hline
$M_H$& 110 & 120 & 130 & 140 & 150 & 160 & 170 & 180\\
\hline\hline
$Z_\gamma^{(g)}$  & 11.4 &  9.9 &  9.4 & 10.1 & 13.5 &      &      &\\
$Z_z^{(g)}$       &      & 23.1 & 12.4 &  8.9 &  7.6 & 13.4 & 18.8 &10.1\\
$Z_w^{(g)}$       &      & 42.1 & 26.0 & 17.0 & 14.8 &  7.0 &  8.0 &16.9\\
$Z_\gamma^{(w)}$  & 13.6 & 12.0 & 11.9 & 13.1 & 16.8 &      &      &\\
$Z_w^{(w)}$       & 16.0 &  7.1 &  4.3 &  3.2 &  3.7 &  2.8 &  2.9  &3.3\\
$Z_\tau^{(w)}$    &  9.1 &  8.8 &  9.9 & 13.0 & 20.7 &      &       &\\
$Z_b^{(t)}$       & 10.5 & 11.4 & 14.3 &      &      &      &      &\\
$Z_\tau^{(t)}$    & 14.1 & 17.0 & 23.3 & 36.7 &      &      &      &\\
$Z_w^{(t)}$       &      &      & 42.0 & 29.0 & 23.0 & 20.0 & 21.0 &24.0\\
$Z_b^{(wH)}$      & 15.0 & 19.0 & 25.0 &      &      &      &      &\\
\hline\hline
\end{tabular}
} 
\FIGURE[t]{
\label{fig:all_channels}
\epsfig{file=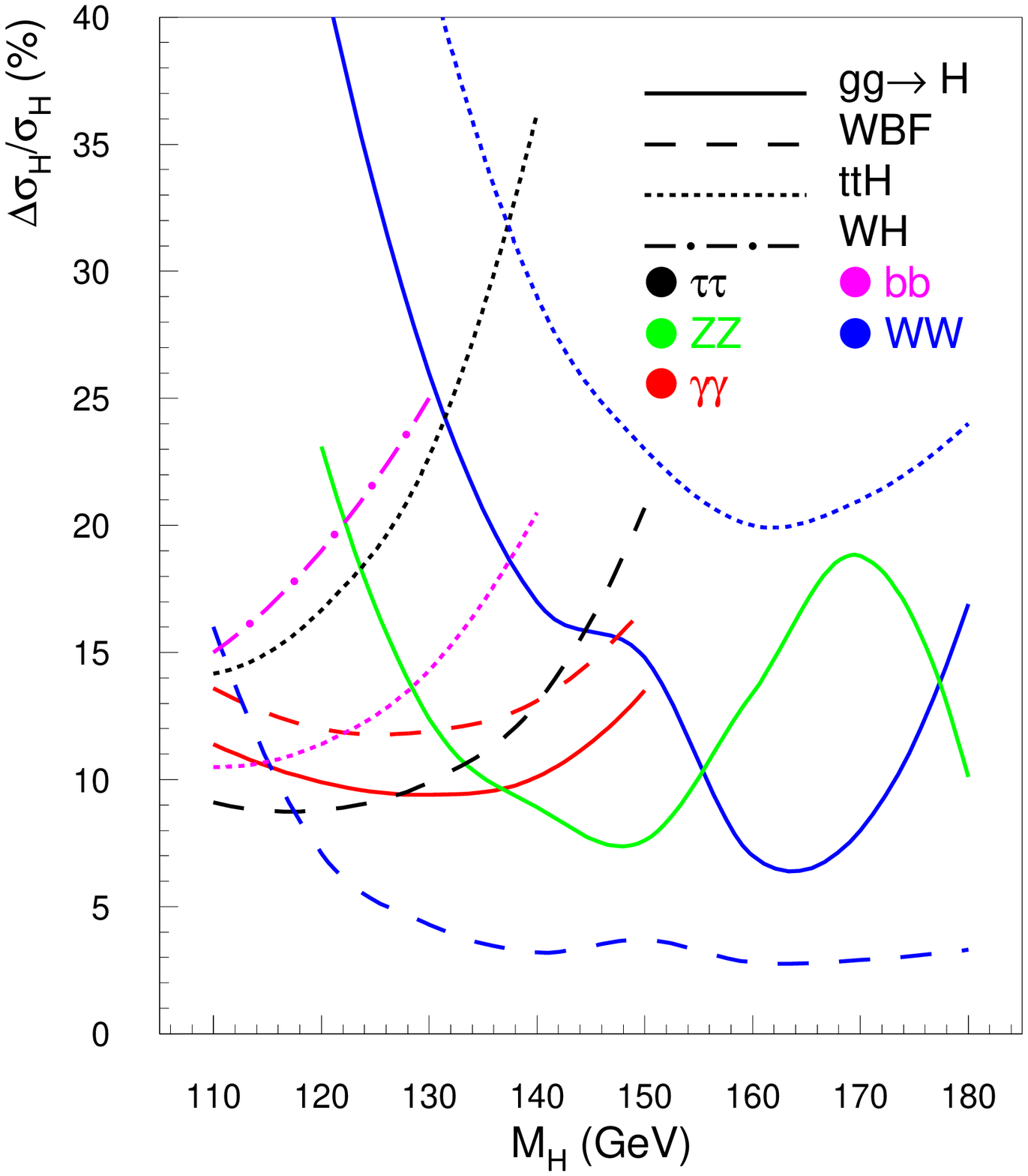,width=15cm}
\caption{
  Relative accuracies on the measurement of the cross section of a
  scalar SM-like Higgs boson in the production+decay channels listed
  in Table~\ref{tab:accuracies}. All channels have been rescaled to a
  total integrated luminosity of 200~fb$^{-1}$ (100~fb$^{-1}$ per
  detector), except $pp\to t\bar{t}H,\,H\to WW$ and $q\bar{q}\to
  WH,\,H\to b\bar{b}$ for which we use
  300~fb$^{-1}$~\cite{Maltoni:2002jr,Drollinger:2002uj}, and $gg\to
  H,\,H\to WW$ that was studied for
  30~fb$^{-1}$~\cite{Zeppenfeld:2000td} (see text).}}

By looking at Table~\ref{tab:accuracies} and
Fig.~\ref{fig:all_channels}, it seems natural to divide the mass range
of an intermediate mass scalar Higgs boson into two main regions,
corresponding to $M_H\!\le\!140$~GeV and $140\!<\!M_H\!\le\!180$~GeV.
All the existing studies belong to one of these two main mass regions.
In particular, in the lower mass region, for $M_H\!\le\!140$~GeV, we
consider the following channels:
\begin{eqnarray}
\label{eq:below_140}
&&gg\rightarrow H\,\,\,
\mbox{with}\,\,\,H\rightarrow\gamma\gamma\,\,\,,\\
&&qq\rightarrow qq H\,\,\,\mbox{with}\,\,\,
H\rightarrow\gamma\gamma,\,\tau\tau,\,WW \,\,\,,\nonumber\\
&&q\bar q,gg\rightarrow t\bar tH\,\,\,\mbox{with}\,\,\,
H\rightarrow b\bar b,\tau\tau\,\,\,,\nonumber
\end{eqnarray}
while in the higher mass region, for $140\!<\!M_H\!\le\!180$~GeV we
consider the following channels:
\begin{eqnarray}
\label{eq:above_140}
&&gg\rightarrow H\,\,\,\mbox{with}\,\,\,H\rightarrow ZZ,\,WW\,\,\,,\\
&&qq\rightarrow qq H\,\,\,\mbox{with}\,\,\,
h\rightarrow WW\,\,\,.\nonumber
\end{eqnarray}
No other convenient choice seems possible at the moment.  Since we
would like to work with a uniform integrated luminosity of
200~fb$^{-1}$ for all channels (100~fb$^{-1}$ per detector), we decide
not to use $gg\to H,\,H\to WW$ for $M_H\!\le\!140$~GeV, since the
corresponding accuracies summarized in Table~\ref{tab:accuracies} have
been obtained using an integrated luminosity of 30~fb$^{-1}$ and are
therefore pretty poor in the lower mass region. Rescaling these
numbers to a much higher luminosity is debatable and we prefer not to
base our analysis on that. On the other hand, we rescale the results
obtained for $pp\to t\bar{t}H,\,H\to b\bar{b}$ from
30~fb$^{-1}$~\cite{Drollinger:2001ym} to 100~fb$^{-1}$, since the
nature of the signal over background shape justifies that.  To account
for both detectors, many results are then rescaled from 100~fb$^{-1}$
to 200~fb$^{-1}$, and this is clearly adequate.

We also note that we could consider $q\bar{q}\to HW,\,H\to b\bar{b}$
instead of $pp\to t\bar{t}H,\, H\to b\bar{b}$. However this does not
seem convenient since, for equal luminosity, the $pp\to t\bar{t}H,\,
H\to b\bar{b}$ channel can be measured with better accuracy and,
together with $pp\to t\bar{t}H,\, H\to\tau^+\tau^-$, allows for a
model independent measurement of $y_b/y_\tau$. Of course, in a very
high luminosity scenario both channels should be combined in order to
obtain better accuracies.  We note that for $M_H\!>\!140$~GeV there is
no channel that can provide a handle on $\Gamma_b$. Both $pp\to
t\bar{t}H,\,H\to b\bar{b}$ and $q\bar{q}\to WH,\,H\to b\bar{b}$ can
possibly be used up to 135-140~GeV, just to cover the entire spectrum
of a light MSSM scalar Higgs boson, but not above 140~GeV. In fact,
for the purpose of our study, we have extrapolated the existing
analysis of $pp\to t\bar{t} H,\,H\to b\bar{b}$
\cite{Drollinger:2001ym} from $M_H\!=\!130$~GeV up to
$M_H\!=\!140$~GeV, for the only purpose of covering the entire MSSM
light scalar Higgs boson mass range.  For higher masses, the
determination of $\Gamma_b$ will remain a problem at hadron colliders.
On the other hand, for the mass region $M_H\!\le\!140$~GeV, certainly
the most interesting one for the scenario we address in this paper, we
can use all channels listed in Eq.~(\ref{eq:below_140}) and we will
see how having the extra possibility of measuring $pp\to
t\bar{t}H,\,H\to\tau^+\tau^-$ is indeed crucial.

We start therefore by discussing the most interesting case of a Higgs
boson with mass $M_H\!\le\!140$~GeV.  This will be indeed the main
focus of our analysis. To be consistent with the picture developed up
to here, we assume that the width of the Higgs boson is mainly
saturated by the decays into $b\bar b$, $\tau^+\tau^-$, $W^+W^-$,
$ZZ$, $gg$, and $\gamma\gamma$.  By using the set of measurements in
Eq.~(\ref{eq:below_140}), we can then solve a system of equations of
the form given in Eq.~(\ref{eq:z_i_j}), one for each channel in
Eq.~(\ref{eq:below_140}). The solution returns the values of the
individual rates $\Gamma_t$, $\Gamma_b$, $\Gamma_\tau$, $\Gamma_w$,
$\Gamma_g$, and $\Gamma_\gamma$ as functions of the
observables $Z^{(i)}_j$ and the total rate $\Gamma$.
The individual $\Gamma_i$ turn out to be:
\begin{eqnarray}
\label{eq:partial_widths}
\Gamma_t&=&\frac{Z^{(t)}_\tau\sqrt{Z^{(w)}_w}}{Z^{(w)}_\tau}\sqrt{\Gamma}\,\,\,,\\
\Gamma_b&=&\frac{Z^{(t)}_bZ^{(w)}_\tau}{Z^{(t)}_\tau\sqrt{Z^{(w)}_w}}\sqrt{\Gamma}
\,\,\,,\nonumber\\
\Gamma_\tau&=&\frac{Z^{(w)}_\tau}{\sqrt{Z^{(w)}_w}}\sqrt{\Gamma}\,\,\,,\nonumber\\
\Gamma_w&=&\sqrt{Z^{(w)}_w}\sqrt{\Gamma}\,\,\,,\nonumber\\
\Gamma_\gamma&=&\frac{Z^{(w)}_\gamma}{\sqrt{Z^{(w)}_w}}\sqrt{\Gamma}
\,\,\,,\nonumber\\
\Gamma_g&=&\frac{Z^{(g)}_\gamma\sqrt{Z^{(w)}_w}}{Z^{(w)}_\gamma}\sqrt{\Gamma}
\,\,\,,\nonumber
\end{eqnarray}
while the total width $\Gamma$ follows from the assumption that
$\Gamma=\Gamma_b+\Gamma_\tau+\Gamma_w+\Gamma_z+\Gamma_g+\Gamma_\gamma$:
\begin{equation}
\label{eq:total_width}
\sqrt{\Gamma}=\frac{1}{\sqrt{Z^{(w)}_w}}\left[
Z^{(w)}_\tau\left(1+\frac{Z^{(t)}_b}{Z^{(t)}_\tau}\right)+
Z^{(w)}_w\left(1+z_{SM}\right)+
\frac{Z^{(w)}_w Z^{(g)}_\gamma}{Z^{(w)}_\gamma}+Z^{(w)}_\gamma 
\right]\,\,\,.
\end{equation}

Furthermore, we note that we can express the decay rates for
$H\to gg$ and $H\to\gamma\gamma$ as the linear combination of terms
due to the SM particles that contribute in the loop, plus an extra
term that accounts for new physics heavy degrees of freedom.
In other words we can write $\Gamma_g$ and $\Gamma_\gamma$ as
\begin{eqnarray}
\label{eq:Gg_Gph}
\Gamma_g&\simeq&A_t^{(g)}\Gamma_t+A_b^{(g)}\Gamma_b+
A_{tb}^{(g)}\sqrt{\Gamma_t\Gamma_b}+\delta_g\\
\Gamma_\gamma&\simeq&A_t^{(\gamma)}\Gamma_t+A_w^{(\gamma)}\Gamma_w+
A_{tw}^{(\gamma)}\sqrt{\Gamma_t\Gamma_w}+\delta_\gamma\nonumber
\end{eqnarray}
where we have neglected all the contributions from SM particles that
are below the level of accuracy expected on the measurement of
$\Gamma_g$ and $\Gamma_\gamma$ (20-30\%), while we have indicated by
$\delta_g$ and $\delta_\gamma$ the unknown loop contributions from new
physics.  Since we have factored the Higgs couplings square into the
corresponding $\Gamma_i$'s, we note that the coefficients
$A_t^{(g,\gamma)}$, $A_b^{(g)}$, $A_{tb}^{(g)}$, $A_w^{(\gamma)}$, and
$A_{tw}^{(\gamma)}$ only depend on $M_H$ and on the mass of the SM
particle in the loop. They represent the explicit contribution of the
top quark, the bottom quark, and the $W$ boson to the $H\to gg$ or
$H\to\gamma\gamma$ loop, and can be taken from the corresponding SM
calculations~\cite{Spira:1998dg}. Using the results in
Eqs.~(\ref{eq:partial_widths}) and (\ref{eq:total_width}), the
expressions of $\delta_g$ and $\delta_\gamma$ follow from
Eq.~(\ref{eq:Gg_Gph}).

\TABLE[htb]{
 \caption{\label{tab:gamma_res} 
   Relative accuracy (\%) for the total width $\Gamma$ and for the
   individual $\Gamma_i$ obtained in the model-independent scenario
   (MI) as well as in the scenario with
   $\Gamma_b/\Gamma_\tau\!=y_{SM}$ (BT) at the LHC with 200 fb$^{-1}$
   total integrated luminosity (100 fb$^{-1}$ per detector). These
   values include the systematic theoretical errors.}
\begin{tabular}{ |c|l||l|l|l|l|l|l|l| }
 \hline
$M_H$ (GeV) &  & 
$\Gamma$ & $\Gamma_t$ & $\Gamma_b$ & $\Gamma_\tau$ & $\Gamma_w$ &
$\Gamma_g$ & $\Gamma_\gamma$\\
\hline\hline
    &&&&&&&&\\
110 & MI & 42.6 & 12.5  & 47.5 & 30.4 & 20.4  & 35.5 & 28.5 \\
    & BT & 25.2 & 14.9  & 25.2 & 25.2 & 11.0  & 31.0 & 22.7 \\ \hline
    &&&&&&&&\\
120 & MI & 38.6 & 12.7  & 46.5 & 26.0 & 19.3  & 33.7 & 23.6 \\
    & BT & 19.4 & 15.7  & 19.9 & 19.9 & 9.6   & 29.3 & 16.7 \\ \hline
    &&&&&&&&\\
130 & MI & 36.0 & 16.0  & 50.8 & 24.6 & 18.2  & 32.7 & 22.0 \\
    & BT & 15.6 & 18.6  & 18.6 & 18.6 & 8.2   & 28.4 & 14.9 \\ \hline
    &&&&&&&&\\
140 & MI & 32.8 & 27.7 & 62.9 & 24.8 & 16.8 & 32.4 & 21.3 \\
    & BT & 12.1 & 25.1  & 19.6 & 19.6 & 7.0 & 28.6 & 14.8 \\ \hline
 \end{tabular}
}

We first determine the relative accuracy on the width, and then use it
to calculate the relative accuracy on the individual rates $\Gamma_i$.
The uncertainty on $\Gamma$, and therefore on the individual rates
$\Gamma_i$, is a complicated function of the uncertainties on the
single experimental channels, that takes into account all theoretical
correlations and relies on the assumption that the single Higgs
branching ratios have a magnitude comparable to the SM Higgs. In
principle, experimental systematics and correlations should also be
taken into account, but this information is not yet available. The
individual production cross sections are also affected by some
theoretical uncertainties, mainly due to higher order radiative
corrections. This can be estimated from existing calculations in terms
of the residual renormalization and factorization scale dependence,
and can be accounted for by adding a systematic error to the
corresponding accuracies in Table~\ref{tab:accuracies}. At the moment
$gg\to H$~\cite{Harlander:2002wh} is affected by the largest
theoretical uncertainty, of the order of 15\%, while the theoretical
uncertainties on $pp\to t\bar tH$ and $qq\to qqH$ are of the order of
$6\%$~\cite{Beenakker:2001rj,Reina:2001sf,Reina:2001bc} and below
$5\%$~\cite{Han:1992hr} respectively. To compare with the analysis
done for the $\Gamma_b/\Gamma_\tau=y_{SM}$ case in
Ref.~\cite{Zeppenfeld:2002ng}, we assume the same conservative
theoretical uncertainties of 20\% for $gg\to H$, of 10\% for $pp\to
t\bar{t}H$, and of 5\% for $qq\to qqH$. Moreover, when comparing with
the $\Gamma_b/\Gamma_\tau=y_{SM}$ case, we will assume an error on
$y_{SM}$ of 7\%, due to the uncertainty on the $b$ quark mass.
\FIGURE[t]{
\vspace*{-1cm}
\caption{\label{fig:gamma_i_accuracies}
  Relative accuracy (\%) on the individual rates $\Gamma_i$ expected
  in the model-independent scenario as well as in a scenario with
  $\Gamma_b/\Gamma_\tau$ fixed to its SM value, at the LHC with
  $\sqrt{s}\!=\!14$~TeV. A total integrated luminosity of 200~fb$^{-1}$
  (100~fb$^{-1}$ per detector) is assumed. The upper plots show the
  accuracies obtained without including any theoretical systematic
  error, while the lower plots show the same accuracies when a
  systematic theoretical error of $20\%$ for the $gg\to H$ channel, of
  $5\%$ for the $qq\to qqH$, and of $10\%$ for $pp\to t\bar{t}H$
  channel are included.}
\hspace*{-1cm}\epsfig{file=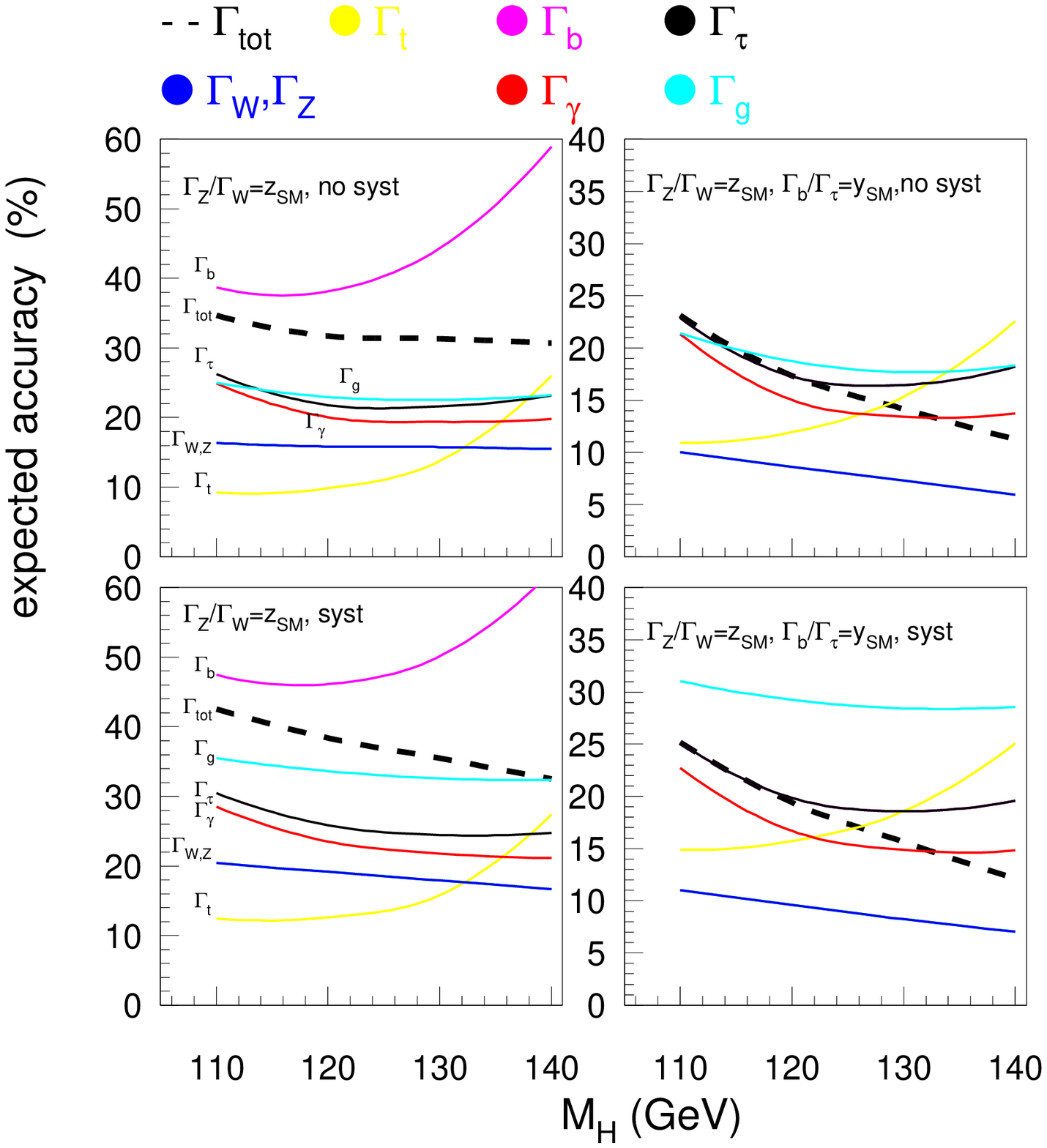,width=16cm,height=17cm} }
Our results for the accuracy on the individual $\Gamma_i$ are reported
in Table~\ref{tab:gamma_res}, for Higgs masses $M_H\!=\!110-140$~GeV,
and are illustrated in Fig.~\ref{fig:gamma_i_accuracies}, both in the
case when the systematic theoretical errors discussed above are
included and when they are not. We prefer to present both results
since some theoretical predictions may be improved in the future, and
the accuracies obtained using the present systematic theoretical
errors can be treated as upper bounds in future analyses.  In the same
figure we also show for comparison the accuracies that we obtain by
following the same approach outlined above, but fixing the ratio
$\Gamma_b/\Gamma_\tau\!=\!y_{SM}$, as done in
Ref.~\cite{Zeppenfeld:2000td,Zeppenfeld:2002ng}. In this case, the
accuracies obtained on the individual $\Gamma_i$ by including the
theoretical uncertainties discussed in this Section show very good
agreement with the results presented in
Ref.~\cite{Zeppenfeld:2002ng,Cavalli:2002vs,Conway:2002kk}.

As expected, by removing some model dependent assumptions the
uncertainty on the single $\Gamma_i$ does not generally improve. It is
however interesting to note that, due to the interplay between
$Z_b^{(t)}$ and $Z_\tau^{(t)}$ in the error propagation analysis, the
accuracy on $\Gamma_t$ is reduced with respect to the model-dependent
case, and is now definitely in the $10-20\%$ range, even after all
theoretical uncertainties has been taken into account. It is also
useful to remember that the actual error on the couplings is half the
error on the corresponding $\Gamma_i$, and therefore, the overall
result of the model independent analysis is very encouraging, with
accuracies on all couplings between 7\% and 25\%. Most of all,
however, we would like to stress the impact of having included the
$pp\to t\bar{t}H,\,H\to\tau^+\tau^-$ channel with two main
considerations.  First of all, we can now determine in a completely
model independent way ratios of couplings like
\begin{equation}
\frac{\Gamma_b}{\Gamma_\tau}=\frac{Z^{(t)}_b}{Z^{(t)}_\tau}\,\,\,\,\,
\mbox{and}\,\,\,\,\,
\frac{\Gamma_t}{\Gamma_g}=\frac{Z^{(t)}_\tau Z^{(w)}_\gamma}
{Z^{(w)}_\tau Z^{(g)}_\gamma}\,\,\,,
\label{eq:ratios_btau_tg}
\end{equation}
with accuracies of the order of 20-30\%, as illustrated in
Fig.~\ref{fig:ratios}. The only residual model dependence in the
determination of the individual $Z^{(i)}_j$ is in the assumption that
the relation between $WWH$ and $ZZH$ weak boson fusion in $qq\to qqH$
is SM-like. However, we note that both ratios in
Eq.~(\ref{eq:ratios_btau_tg}) do not actually depend on such
assumption. In fact, only $\Gamma_t/\Gamma_g$ depends on $Z^{(w)}_j$
observables, but the model dependence in assuming
$\Gamma_z/\Gamma_w\!=\!z_{SM}$ cancels between numerator and
denominator.

Furthermore, we note that, by adding the $pp\to
Ht\bar{t},\,H\to\tau^+\tau^-$ channel, we provide the possibility of
testing the $\Gamma_z/\Gamma_w\!=\!z_{SM}$ universality assumption
even for $M_H\!<\!140$~GeV (!) by comparing the two ratios
\begin{equation}
\label{eq:zw_universality}
\frac{Z^{(wH)}_b}{Z^{(t)}_b} \,\,\,\,\,\,\mbox{and}\,\,\,\,\,\,
\frac{Z^{(w)}_\tau}{Z^{(t)}_\tau}\,\,\,,
\end{equation}
at high luminosity of the order of 300~fb$^{-1}$. Indeed, while the
first ratio is always proportional to $\Gamma_w/\Gamma_t$, the second
ratio is proportional to a combination of $\Gamma_w/\Gamma_t$ and
$\Gamma_z/\Gamma_t$. The comparison between the two ratios can
therefore test the assumption that $\Gamma_z/\Gamma_w\!=\!z_{SM}$.
Using the accuracies listed in Table~\ref{tab:accuracies}, both ratios
in Eq.~(\ref{eq:zw_universality}) can be measured at the 20-30\% level
depending on the value of $M_H$.
\FIGURE[t]{
\label{fig:ratios}
\epsfig{file=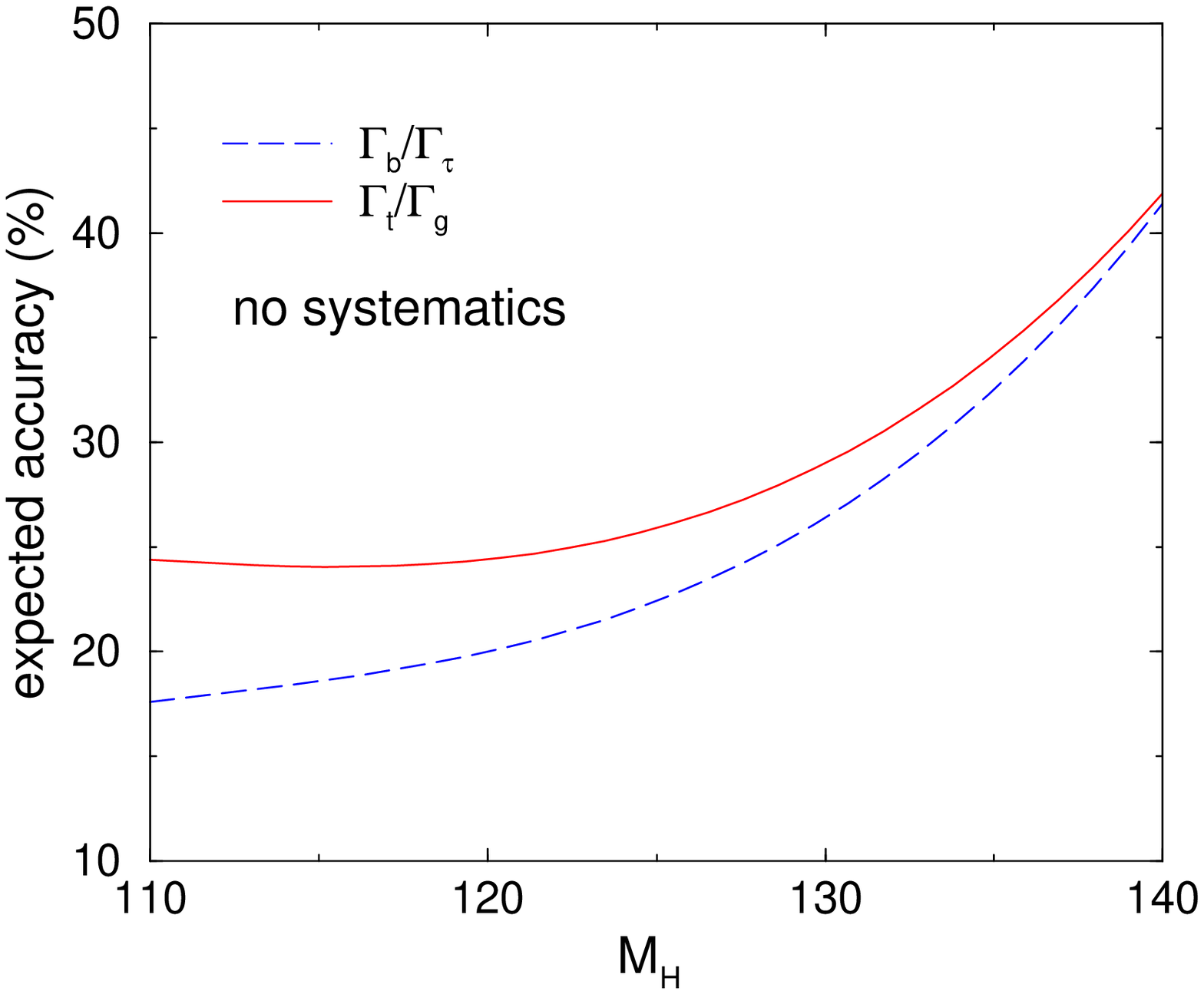,width=10cm,height=8cm}
\epsfig{file=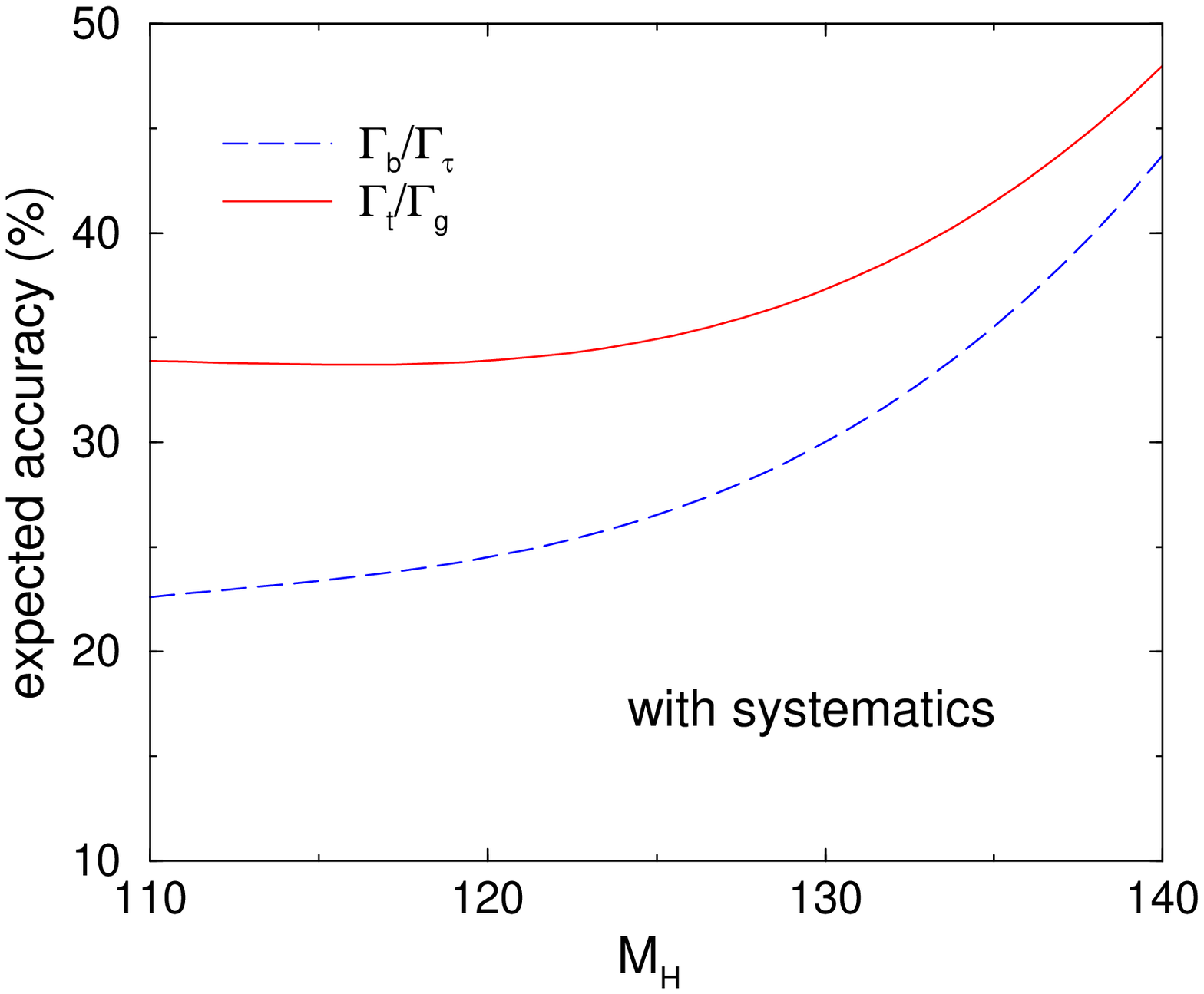,width=10cm,height=8cm}
\caption{\label{fig:ratios_accuracies} 
  Relative accuracy (\%) on the measurement of the ratios
  $\Gamma_b/\Gamma_\tau$ and $\Gamma_t/\Gamma_g$ without (top) and
  with (bottom) systematic theoretical errors.}  }

Finally, we have studied the accuracy with which $\delta_g$ and
$\delta_\gamma$ can be determined. We summarize our results in
Fig.~\ref{fig:ratios_accuracies} where we plot the expected accuracy
for different values of $\delta_g/\Gamma_g$ and
$\delta_\gamma/\Gamma_\gamma$, and $M_H\!=\!120$~GeV. According to the
definition in Eq.~(\ref{eq:Gg_Gph}), $\delta_g$ and $\delta_\gamma$
represent contributions to the $Hgg$ and $H\gamma\gamma$ vertices that
are not already accounted for in deviations of the Higgs couplings to
the SM degrees of freedom.  Therefore they account for loop effects
from non-SM heavy degrees of freedom that do not affect the Higgs
width directly.  It is interesting to note from
Fig.~\ref{fig:deltas_accuracies} that only deviations of the order of
50\% or bigger will be measurable with sufficient accuracy. The curves
plotted in Fig.~\ref{fig:deltas_accuracies} are obtained by allowing
extra large $\delta_g$ or $\delta_\gamma$ contributions in either
$\Gamma_g$ or $\Gamma_\gamma$, i.e. not in both at the same time.
Therefore this plot contains some model dependence, and has to be
taken just as an indication of the fact that it will be hard to
measure small purely loop effects at the LHC through $\Gamma_g$ and
$\Gamma_\gamma$.

We conclude by examining the mass region
$140\!\le\!M_H\!\le\!180$~GeV, where we consider the channels listed
in Eq.~(\ref{eq:above_140}).  Unfortunately, unless a very exotic
Higgs boson is discovered, case that is not considered in this
analysis, no model independent way to determine the couplings of a
scalar Higgs to both SM fermions and gauge bosons can be developed.
In fact, none of the channels listed in Eq.~(\ref{eq:above_140}) will
allow a measurement of the scalar Higgs boson couplings to the $\tau$
lepton or to the bottom quark.  The decays into the SM gauge bosons
dominate and, as mentioned before, this offers the important
possibility to precisely test the $\Gamma_z/\Gamma_w\!=\!z_{SM}$
assumption at the 20-30\%
level~\cite{Zeppenfeld:2000td,Zeppenfeld:2002ng}.  Moreover, since
$pp\to t\bar{t}H,\, H\to WW$ becomes available, the ratio
$\Gamma_t/\Gamma_g$ can be tested in a model independent way through a
measurement of $Z^{(t)}_w/Z^{(g)}_w$, although this will require
luminosities of the order of 300~fb$^{-1}$ for the measurement of
$Z^{(t)}_w$ \cite{Maltoni:2002jr}.  Since the $pp\to
t\bar{t}H,\,H\to\tau^+\tau^-$ channel that we have studied in this
paper cannot be measured with good accuracy for $M_H\!>\!140$~GeV, in
this mass region our analysis follows the pattern of existing studies
\cite{Zeppenfeld:2000td,Zeppenfeld:2002ng,Maltoni:2002jr} to which we
refer for more details.

\FIGURE[t]{
\noindent
\epsfig{file=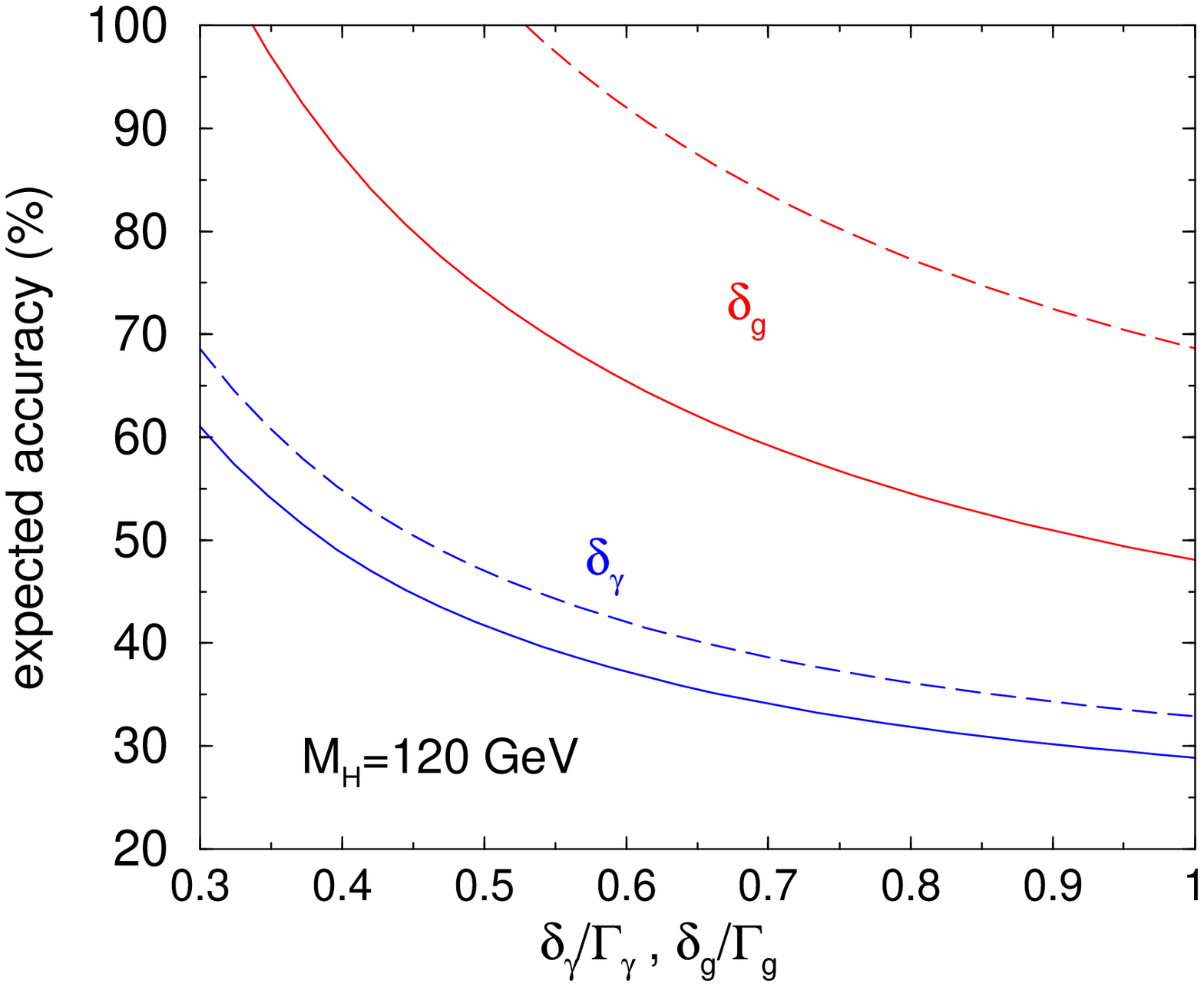,width=10cm,height=8cm}
\caption{\label{fig:deltas_accuracies} 
  Relative accuracies (\%) for $\delta_g$ and $\delta_\gamma$ obtained
  in the model-independent scenario, for a scalar Higgs boson with
  $M_H\!=\!120$~GeV, at the LHC with $\sqrt{s}\!=\!14$~TeV.  We
  illustrate both the case in which systematic theoretical errors have
  been (dashed) and have not been (solid) included.  A total
  integrated luminosity of 200 fb$^{-1}$ (100 fb$^{-1}$ per detector)
  is assumed.  }  }

\section{Conclusions}
\label{sec:conclusions}

In this paper we have studied the potential of the LHC to observe a
relatively light scalar Higgs boson in the $pp\to t\bar{t}H,\,H\to
\tau^+\tau^-$ process at $\sqrt{s}\!=\!14$~TeV.  The study has been
mainly done at the parton level, but taking into account detector
energy resolution and with the correct simulation of $\tau$-leptons
decays. Following the strategy outlined in Section~\ref{sec:tautau},
we have shown that for 100~fb$^{-1}$ of total integrated luminosity
one has 34-8 signal events for $M_H\!=\!110-140$~GeV respectively,
against about 12 background events.

We have shown that one can use the $pp\to t\bar{t}H,\,H\to
\tau^+\tau^-$ process to measure the $y_t\times y_\tau$ product of
Yukawa couplings. This is a crucial point, since the addition of the
$pp\to t\bar{t}H,\,H\to\tau^+\tau^-$ channel to other already studied
Higgs production channels allows to measure the $y_\tau$ and $y_b$
Yukawa couplings {\it independently} and removes the assumption of
$y_b/y_\tau$ universality used in all previous studies. This allows to
be sensitive to non-trivial radiative corrections breaking the
$y_b/y_\tau$ universality in models of new physics beyond the SM,
notably Supersymmetric models.

For the case of a light scalar Higgs boson, with mass $M_H\!\le\!
140$~GeV, we have shown how to derive the accuracies on the
measurement of the width and of the individual $y_t$, $y_b$, $y_\tau$,
$y_w$, $y_g$, and $y_\gamma$ couplings to SM fermions and gauge
bosons.  As an example, we have presented the numerical values of
these accuracies for a SM-like scenario. Results are encouraging, even
when all known systematics are taken into account. Accuracies of the
order of 10-20\% are expected for most couplings. In this scenario,
the ratios of $y_b/y_\tau$ and $y_t/y_g$ can be measured in a model
independent way, with accuracies that also fall in the 15-20\% range.
In addition, adding $pp\to Ht\bar{t},\,H\to\tau^+\tau^-$ provides
the possibility of testing the $y_z/y_w$ universality for Higgs boson
masses $M_H\!<\!140$~GeV.

We have also investigated the sensitivity to non SM particles
contributing to the $Hgg$ or $H\gamma\gamma$ loop-induced vertices,
and determined that only very large deviations, of the order of 50\%
or more, will be measured with sufficient accuracy.

Our final results, presented in Table~\ref{tab:gamma_res} and
illustrated in
Figs.~\ref{fig:gamma_i_accuracies}-\ref{fig:deltas_accuracies}, show
that a study of the $pp\to t\bar{t}H,\,H\to\tau^+\tau^-$ process
allows us to make an important step towards a model independent
measurement of the couplings of a light scalar Higgs boson at the LHC.

\acknowledgments

We thank Dieter Zeppenfeld and Horst Wahl for very informative
discussions. This work is supported in part by the U.S. Department of
Energy under grant DE-FG02-97ER41022.

\bibliography{higgs_couplings}
\end{document}